\documentclass[fleqn,12pt,draft]{article} 
\usepackage[intlimits,reqno]{amsmath}
\def\oo{\infty}
\def\Z#1{\zeta(#1)}
\def\Li{{\mathrm{Li}}}

\def\NC{{\it Nuovo Cimento }}
\def\NuPh{{\it Nucl. Phys. }}
\def\PL{{\it Phys. Lett. }}
\def\PR{{\it Phys. Rev. }}

\def\IJMP{{\it Int. J. Mod. Phys. }}

\def\IOMOG{I^{HO}}
\def\INOMOG{I^{NH}}

\def\e{\epsilon}

\def\SYS{{\tt SYS}}
\def\TITLE{High-precision $\e$-expansions of massive four-loop vacuum bubbles}
\def\IDENTIFY{\par (S. Laporta, \TITLE)}
\def\CAPTIONFIGA{Four-loop vacuum bubble diagrams.}
\def\CAPTIONFIGB{Three-loop self-mass diagram.}

\def\eqref#1{Eq.(\ref{#1})}
\def\eqrefb#1#2{Eqs.(\ref{#1})-(\ref{#2})}
\def\itref#1{(\ref{#1})}

\def\FigureOne
{
\begin{figure}
\begin{center}
   \begin{picture}(370,180)(0,0)
   \thicklines
   \put(040,100){\makebox(0,0)[lb]{
   \put(000,000){\circle{40}}
   \put(-020,000){\line(1,0){40}}
   \qbezier(-020,000)(000,+020)(19.5,000)
   \qbezier(-020,000)(000,-020)(19.5,000)
   \put(-016,021){\text{\footnotesize{$k_1$}}}   
   \put(-018,-023){\text{\footnotesize{$k_2$}}}   
   \put(-015,-040){\text{\large $V_1\ V_2$}}   
   }}

   \put(110,100){\makebox(0,0)[lb]{
   \put(000,000){\circle{40}}
   \qbezier(-020,000)(-010,+010)(000,+020)
   \qbezier(-020,000)(-003,+003)(000,+020)
   \qbezier(-020,000)(-003,-003)(000,-020)
   \put(-008,-040){\text{\large $V_3$}}   
   }}
   
   \put(180,100){\makebox(0,0)[lb]{
   \put(000,000){\circle{40}}
   \qbezier(-017.3,-010)(-005.2,+003)(000,020)
   \qbezier(+017.0,-010)(+005.2,+003)(000,020)
   \qbezier(-017.3,-010)(000,-006)(+017.0,-010)
   \put(-008,-040){\text{\large $V_4$}}   
   }}
   \put(250,100){\makebox(0,0)[lb]{
   \put(000,000){\circle{40}}
   \qbezier(000,000)(+010,+010)(000,020)
   \qbezier(000,000)(-010,+010)(000,020)
   \put(000,000){\line(+5,-3){17}}
   \put(000,000){\line(-5,-3){17}}
   \put(-024,+016){\text{\footnotesize{$k_1$}}}
   \put(+016,+016){\text{\footnotesize{$k_2$}}}
   \put(-020,-001){\vector(0,1){3}}
   \put(+020,-001){\vector(0,1){3}}
   \put(-015,-040){\text{\large $V_5\ V_6$}}   
   }}
   \put(320,100){\makebox(0,0)[lb]{
   \put(000,000){\circle{40}}
   \qbezier(-020,000)(-003,+003)(000,+020)
   \qbezier(-020,000)(-003,-003)(000,-020)
   \put(-020,000){\line(1,0){40}}
   \put(-008,-040){\text{\large $V_7$}}   
   }}
   \put(040,010){\makebox(0,0)[lb]{
   \put(000,000){\circle{40}}
   \qbezier(-014.1,-014.1)(-006,000)(-014.1,014.1)
   \qbezier(+014.0,-014.1)(+006,000)(+014.0,014.1)
   \qbezier(-014.1,-014.1)(000,-006)(+014.0,-014.1)
   \put(-010,-040){\text{\large $V_8$}}   
   }}
   \put(110,010){\makebox(0,0)[lb]{
   \put(000,000){\circle{40}}
   \put(000,015){\circle{10}}
   \put(000,000){\line(0,1){10}}
   \put(000,000){\line(+5,-3){17}}
   \put(000,000){\line(-5,-3){17}}
   \put(-008,-040){\text{\large $V_9$}}   
   }}
   \put(180,010){\makebox(0,0)[lb]{
   \put(000,000){\circle{40}}
   \put(-020,000){\line(1,0){40}}
   \put(000,-020){\line(0,1){40}}
   \put(-024,+016){\text{\footnotesize{$k_1$}}}
   \put(+015,-020){\text{\footnotesize{$k_2$}}}
   \put(-016.37,+012){\vector(1,2){0}}
   \put(+016.37,-012){\vector(-1,-2){0}}
   \put(-020,-040){\text{\large $V_{10}\ V_{11}$}}   
   }}
   \put(250,010){\makebox(0,0)[lb]{
   \put(000,000){\circle{40}}
   \put(-010,-017.3){\line(0,+1){34.6}}
   \put(+010,-017.3){\line(0,+1){34.6}}
   \put(-010,000){\line(1,0){20}}
   \put(-010,-040){\text{\large $V_{12}$}}   
   }}
   \put(320,010){\makebox(0,0)[lb]{
   \put(000,000){\circle{40}}
   \put(-014,+014){\line(+1,-1){28}}
   \qbezier(-014,-014)(-008,-008)(-002,-002)
   \qbezier(+014,+014)(+008,+008)(+002,+002)
   \qbezier(-020,000)(-012,-000)(-004,000)
   \qbezier(+020,000)(+012,-000)(+004,000)
   \put(-010,-040){\text{\large $V_{13}$}}   
   }}
   \end{picture}
\end{center}
\phantom{ }\vspace{3truecm}\phantom{ } 
\caption{}
 \label{figureone}
 \end{figure}
}
\def\FigureTwo
{
\begin{figure}
\begin{center}
   \begin{picture}(150,100)(0,0)
   \thicklines
   \put(075,010){\makebox(0,0)[lb]{
   \put(-040,000){\line(-1,0){15}}
   \put(+040,000){\line(+1,0){15}}
   \qbezier(+000.0,+040.0)(-022.7,+039.3)(-034.6,+020.0)
   \qbezier(-034.6,+020.0)(-045.4,+000.0)(-034.6,-020.0)
   \qbezier(-034.6,-020.0)(-022.7,-039.3)(+000.0,-040.0)
   \qbezier(+000.0,-040.0)(+022.7,-039.3)(+034.6,-020.0)
   \qbezier(+034.6,-020.0)(+045.4,+000.0)(+034.6,+020.0)
   \qbezier(+034.6,+020.0)(+022.7,+039.3)(+000.0,+040.0)
   \qbezier(-040.0,+000.0)(+000.0,+040.0)(+040.0,+000.0)
   \qbezier(-040.0,+000.0)(+000.0,-040.0)(+040.0,+000.0)
   }}
   \end{picture}
\end{center}
\phantom{ }\vspace{3truecm}\phantom{ } 
\caption{}
 \label{figuretwo}
 \end{figure}
}
\newcommand\mytoday{\number\day\space \ifcase\month\or
  January\or February\or March\or April\or May\or June\or
    July\or August\or September\or October\or November\or December\fi
      \space\number\year}

\begin{document}
%%%%%%%%%%%%%%%%%%%%%%%%%%%%%%%%%%%%%%%%%%
%%%%%%%%%%%%%%%%%%%%%%%%%%%%%%%%%%%%%%%%%%
% for `article'
\title{\vspace{1cm} \TITLE }
\author{S. Laporta\thanks{{E-mail: \tt laporta{\char"40}bo.infn.it}} \\ 
 \hfil \\ {\small \it Dipartimento di Fisica, Universit\`a di Bologna, }
 \hfil \\ {\small \it Via Irnerio 46, I-40126 Bologna, Italy} 
 }
\date{}
%%%%%%%%%%%%%%%%%%%%%%%%%%%%%%%%%%%%%%%%%%
%%%%%%%%%%%%%%%%%%%%%%%%%%%%%%%%%%%%%%%%%%
\maketitle
\vspace{-7.5cm} \hspace{12.5cm} {\mytoday} \vspace{+7.5cm}
\vspace{1cm}
\begin{abstract}
In this paper we calculate at high-precision the expansions in $\e=(4-D)/2$ 
of the master integrals of 4-loop vacuum bubble diagrams with equal masses, 
using a method based on the solution of systems of difference equations. 
We also show that the analytical expression of a related on-shell 3-loop 
self-mass master integral contains
new transcendental constants made up of complete elliptic integrals of 
first and second kind.
\end{abstract}

\vspace{1.5cm}
PACS number(s):
\par  12.20.Ds, 12.38.Bx
\par
Keywords:  Feynman diagram, master integral, difference equation.
  
\pagenumbering{roman}
\setcounter{page}{0}
\vfill\eject 
\pagenumbering{arabic}
\setcounter{page}{1}
%%%%%%%%%%%%%%%%%%%%%%%%%%%%%%%%%%%%%%%%%%%%%%%%%%%%%%%%%%%%%%%%%%%%%%%%%%%%%%
%\tableofcontents %% amsart and article
%\listoffigures  %% only article
%\listoftables  %% only article
%\section{Introduction}
One popular approach to calculating Feynman diagrams relies on the use of 
integration-by-parts identities\cite{Tkachov,Tkachov2}. 
By using algebraic identities the contribution of a diagram is reduced 
to a combination of irreducible integrals,
the so-called master integrals.
Only for these an explicit calculation is needed.
For diagrams containing different masses and momenta, 
it is possible to express a multi-scale master integral in terms of 
single-scale integrals by means of well-known technique of 
asymptotic expansions\cite{Asy_eucl1,Asy_eucl2,Asy_eucl3,Asy_eucl4,Asy_reg}.
This technique has had numerous applications in the last years; we refer 
the reader to the reviews \cite{Asy_summ,Harlander,Steinhauser} 
and the references therein.

The single-scale master integrals contain at most 
two possible values of masses, $0$ and $m$, and external lines on-shell;  
being pure numbers, they can be calculated analytically or numerically
at high precision once and for all.
Some results are already available in the literature:
two-loop single-scale self-mass master integrals and
three-loop vacuum master integrals were evaluated analytically 
respectively in Refs.\cite{Single1,Single2,Single3,Single4}
and Ref.\cite{2e}, for all possible combinations of masses $0$ and $m$;
among the numerous applications, for example, 
three-loop vacuum integrals were used for calculating the
relation\cite{MS} between the ${\overline{\rm MS}}$ and the on-shell 
quark mass at order $\alpha_s^3$.

The three-loop single-scale self-mass master integrals appearing 
in the 3-loop QED contribution
to the electron $g$-$2$ were calculated in analytical form in
Ref.\cite{3-loop,polon};
other three-loop self-mass master integrals appearing in QCD diagrams 
were calculated analytically in Ref.\cite{Ritbergen}.
An analytical calculation of self-mass master integrals which appear
in the 4-loop QED 
contribution to the electron $g$-$2$ seemed not to be feasible 
with the techniques used so far, 
so that the author has recently developed a new 
method of calculation of master integrals based on the high-precision numerical
solution of systems of difference equations \cite{Lapdif1,Lapdif2}. 
As first applications of this method, the three-loop all-massive self-mass  
master integrals and deeper expansions of 3-loop QED master integrals 
have been calculated at high precision\cite{Lapdif1,Lapdif2,Lapdif3}. 

In this paper we apply this method for the first
time at 4-loop level, by calculating \emph{all} 
the 4-loop vacuum-bubble master integrals with all massive lines. 
We think the results obtained are of some importance in view of the 
applications
to 4-loop $g$-$2$, since the values of the these vacuum integrals can be used
 as initial conditions
for the integration of differential equations in masses\cite{ko1}
 and external momenta\cite{re1} 
in order to work out the 4-loop $g$-$2$ self-mass master integrals.

We consider here only the master integrals which do not factorize into 
a product of integrals with fewer loops, already known.
There are 10 different topologies for these integrals,
shown in Fig.1, of which that corresponding to $V_{13}$ is not planar. 
Three diagrams have two master integrals.
The Laurent expansion in  $\e=(4-D)/2$ of the master integrals are: 

%%%%%%%%%%%%%%%%%%%%%%%%%%%%%%%%%%%%%%%%%%%%%%%%%%%%%%%%%%%%%%%%%%%%%%%%%%%%%%
%%\FigureOne
%%%%%%%%%%%%%%%%%%%%%%%%%%%%%%%%%%%%%%%%%%%%%%%%%%%%%%%%%%%%%%%%%%%%%%%%%%%%%%
%.........................................................................
\begin{multline}
\label{V1}
%
%(a)
V_{1}=
           -2.5\e^{-4}
           -11.6666666666666666666666666667\e^{-3}
\\	   
           -31.7013888888888888888888888889\e^{-2}
           -67.5289351851851851851851851852\e^{-1}
\\	   
           -140.2205432875405077693377105
           -573.5347004606566057988634367\e
\\	   
           -2756.21982203281444579061754077\e^2
           -18239.9256745938582375474975495\e^3
\\	   
           -86167.478580009225683821281651\e^4
           -468163.766003582350760458083774\e^5
\\	   
           -1976611.11224497547116753862679\e^6
           -9573236.83728658266329080404209\e^7
	   +O(\e^8)\ ,
\end{multline}
%.........................................................................
\begin{multline}
%a2 (k1.k2)^2
V_{2}=
         -1.6875\e^{-4}
         -7.8125\e^{-3}
         -21.2096354166666666666666666667\e^{-2}
\\	   
         -44.7695529513888888888888888889\e^{-1}
         -97.0765348083978788434432266577
\\	   
         -290.923474358239921908941634044\e
         -1719.80956172642828117680003592\e^2
\\	   
         -8934.7307852874907107364237238\e^3
         -51529.1688840498204793229332431\e^4
\\	   
         -236942.747528861416676127442283\e^5
         -1169551.39766202377654297049171\e^6
\\	   
         -4923196.24852687778836136788114\e^7
	 +O(\e^8)
\end{multline}
%.........................................................................
\begin{multline}
%
%(b)
V_{3}=
            1.625\e^{-4}
           +10.2291666666666666666666666667\e^{-3}
\\	   
           +33.4657920462820508294798181834\e^{-2}
           +83.6384721220072849057411859385\e^{-1}
\\	   
           +142.482921255630165397059484622
           +73.8226346988098680135859408617\e
\\	   
           -2215.06968731839973043724485006\e^2
           -10483.810544518473050561125786\e^3
\\	   
           -75868.1741728976518970530475501\e^4
           -266235.221240550768248796525565\e^5
\\	   
           -1562163.76406718387868426308061\e^6
           -5089761.51470684069337522895124\e^7
	   +O(\e^8)\ ,
\end{multline}
%.........................................................................
\begin{multline}\label{V4}
%
%(c)
V_{4}=
            1.5\e^{-4}
           +9.5\e^{-3}
           +33.5\e^{-2}
           +59.8938292905212171438007855155\e^{-1}
\\	   
           -6.77093494521947617804741496782
           -927.184486122109949342449134087\e
\\	   
           -7607.58935727331640672412576731\e^2
           -36043.9315389277686247446140344\e^3
\\	   
           -199692.43129042192811514839511\e^4
           -784516.116603849522465732718633\e^5
\\	   
           -3938768.43031938169221908816973\e^6
           -14490248.5601255703505458679593\e^7
	   +O(\e^8)\ ,
\end{multline}
%.........................................................................
\begin{multline}
%
%(d)
V_{5}=
           -0.25\e^{-4}
           -2.33333333333333333333333333333\e^{-3}
\\	   
           -11.7695293784587392828307280615\e^{-2}
           -47.3155150442894117036246868531\e^{-1}
\\	   
           -139.695412330272368658893294367
           -952.24968928016088147255137961\e
\\	   
           -1559.77276310374672922449535587\e^2
           -17433.2756155912579714527203251\e^3
\\	   
           -18039.431294353693419505748424\e^4
           -300299.661276324759969342746297\e^5
\\	   
           -223362.891943298104957627580543\e^6
           -5002799.01434332803606709511202\e^7
	   +O(\e^8)\ ,
\end{multline}
%.........................................................................
\begin{multline}
%
%(d1) (k1+k2)^2
V_{6}=
            2.375\e^{-4}
           +15.9791666666666666666666666667\e^{-3}
\\	   
           +55.370266375339080107172526045\e^{-2}
           +180.944159673951583547109549717\e^{-1}
\\	   
           +496.624640289958679240601686338
           +2103.52942297240529902308231413\e
\\	   
           +4385.91019392343391919235019438\e^2
           +29880.7052368668975066622604109\e^3
\\	   
           +45750.0902911026047573732270476\e^4
           +472601.136971716877928457965404\e^5
\\	   
           +573888.829426582538928155340449\e^6
           +7686577.54749127119979436449377\e^7
	   +O(\e^8)\ ,
\end{multline}
%.........................................................................
\begin{multline}
%
%(e)
V_{7}=
           -0.5\e^{-4}
           -4.41666666666666666666666666667\e^{-3}
\\	   
           -18.4682782839316233281953131334\e^{-2}
           -68.0334660202028428854058007578\e^{-1}
\\	   
           -295.653297261799619866589394558
           -1111.65699468499496145662198062\e
\\	   
           -4624.23997399795582698262827805\e^2
           -18166.6470161776485664443287094\e^3
\\	   
           -72747.4433773622788330319854055\e^4
           -294544.59963387669806853126121\e^5
\\	   
           -1152092.06523577988360701658787\e^6
           -4748621.41279753209699735445982\e^7
	   +O(\e^8)\ ,
\end{multline}
%.........................................................................
\begin{multline}
%
%(f)
V_{8}=
           -0.75\e^{-4}
           -6.25\e^{-3}
           -23.7024174258974349922929697001\e^{-2}
\\	   
           -83.9541325752545325454615616904\e^{-1}
           -438.152668276175094650239945304
\\	   
           -1151.7188201038655941347197846\e
           -7624.47573539137683288770528724\e^2
\\	   
           -16496.7827202671608148012112968\e^3
           -127976.07440527009995741832362\e^4
\\	   
           -245588.151837894378828676455499\e^5
           -2103545.82723681108089748244607\e^6
\\	   
           -3759879.03947407762892613884153\e^7
	   +O(\e^8)\ ,
\end{multline}
%.........................................................................
\begin{multline}
%
%(g)
V_{9}=
            1.80308535473939142809960724227\e^{-2}
           -9.0439355503127513874520962153\e^{-1}
\\	   
           +42.0771149080147860090872289595
           -185.506435563765087317986949292\e
\\	   
           +794.690451268783109449998200836\e^2
           -3340.85301017420853074405241137\e^3
\\	   
           +13860.9560593967975329182060014\e^4
           -56963.5538585614643909090940607\e^5
\\	   
           +232474.430355572470669990540268\e^6
           -943890.666478469607638755776088\e^7
	   +O(\e^8)\ ,
\end{multline}
%.........................................................................
\begin{multline}
%
%(h)
V_{10}=
            5.18463877571684963165682743229\e^{-1}
           -38.9591108409575465929423742741
\\	   
           +216.540829715907801334069893492\e
           -1056.20279561819887173122551098\e^2
\\	   
           +4804.63905121999951090192731115\e^3
           -20965.3713155251719483741628451\e^4
\\	   
           +89093.2196878747322139561202477\e^5
           -372004.961361167636317053811767\e^6
\\	   
           +1534700.00918085457184730424284\e^7
	   +O(\e^8)\ ,
\end{multline}
%.........................................................................
\begin{multline}
%
%(h1) k1.k2
V_{11}=
           -1.36872356982626095206640482818\e^{-2}
           -10.1337693663646787142726963524\e^{-1}
\\	   
           +25.6153496410310292617132790055
           -334.538889614391947948784653628\e
\\	   
           +984.338776062634546439769565907\e^2
           -7183.20521761939021795215293201\e^3
\\	   
           +21442.9895505509641128293112066\e^4
           -132391.959284919922968209921626\e^5
\\	   
           +396184.588583865634141091817143\e^6
           -2278903.80544414662394641296011\e^7
	   +O(\e^8)\ ,
\end{multline}
%.........................................................................
\begin{multline}
%
%(i)
V_{12}=
            1.34894802170970895986445430292
           -8.34992524160927013803891299755\e
\\	   
           +40.4425656120211378353957075462\e^2
           -177.294826919557277843981859823\e^3
\\	   
           +742.428926688827707784575315859\e^4
           -3039.49399009410599098768805594\e^5
\\	   
           +12302.4392626409424372470869128\e^6
           -49506.0256176923061393145333901\e^7
	   +O(\e^8)\ ,
\end{multline}
%.........................................................................
\begin{multline}
\label{V13}
%
%(j)
V_{13}=
            0.997672576874263051049093586736
           -5.83175720330362704111639113258\e
\\	   
           +27.5586906484187171474221359635\e^2
           -119.248620281553405809848697481\e^3
\\	   
           +495.915629767162238727475039832\e^4
           -2022.83520871375744081923368263\e^5
\\	   
           +8171.6838591733877817993555444\e^6
           -32850.412079699027325130879249\e^7
	   +O(\e^8)\ .
\end{multline}
%.........................................................................
The master integrals $V_j$ are defined as
\begin{equation*}
V_j=
  \left({\pi^{D/2}\Gamma(1+\epsilon)}\right)^{-4}
  \int{d^D k_1 \; d^D k_2 \; d^D k_3 \; d^D k_4}\ \dfrac{P_j}{Q_j} \ ,
\end{equation*}
where $P_2=(k_1 \cdot k_2)^2$, $P_6=(k_1+k_2)^2$, $P_{11}=k_1\cdot k_2$ and
all other $P_j=1$; 
$Q_j$ is the product of denominators of the corresponding 
$j$th diagram of Fig.1, each with unit mass.
The expressions of the finite integrals $V_{12}$ and $V_{13}$ have been checked
by comparing them with values obtained by performing Monte-Carlo integrations
over Feynman parameters.

The rapid growth of the coefficients is due to the fact that the integrals,
seen as functions of $D$, have poles near $D=4$. 
For example, in the case of $V_{13}$ the nearest poles
are $D=9/2, 5, 16/3$, and $11/2$; 
by factorizing out these poles, one obtains a series with decreasing coefficients:
\begin{multline}
V_{13}= 
\left[
(1+4\e)
(1+2\e)
\left(1+\dfrac{3}{2}{\e}\right)
\left(1+\dfrac{4}{3}{\e}\right)
\right]^{-1}
\biggl(
  .9976726 
 +2.981017 \e 
 +2.981995 \e^2 \\
 +1.315019 \e^3  
 +.432643  \e^4 
 +.0999339 \e^5 
 +.0150665 \e^6 
 +.0086152 \e^7 
 +O(\e^8)
\biggr)\ .
\end{multline}
%with coefficients decreasing.

Now we sketch here the method used for obtaining \eqrefb{V1}{V13};
for a complete description the reader is referred to Refs.\cite{Lapdif1,Lapdif2}.
For each master integral $V_j$ we have chosen a denominator and 
raised it to $n$, and we have built a system of difference equations
between the functions $V_j(n)$. 
The difference equation for a given master integral contains in the 
r.h.s. only master integrals with less denominators,
which are simpler. The equations of the system are solved one at once,
 beginning with that corresponding to the simplest master integral,
$V_1$, and ending with that corresponding to the most complex,
$V_{13}$. The necessary boundary conditions at $n\to\oo$ 
are expressed by 3-loop integrals
which are calculated by solving other systems of difference equations.
The homogeneous and nonhomogeneous solutions of the difference equations 
are expanded in factorial series or alternatively transformed into integral 
representations, and evaluated numerically for suitable integer values
of the exponents. By taking these results as input values and  
by using the equations of the system as recurrence relations, one 
recovers the values of the master integrals $V_j(1)$, \eqrefb{V1}{V13}.

For example, the function $V_1(n)$ satisfies the fourth-order 
difference equation
\begin{equation}\label{eqv1}
p_1 V_1(n+1) +p_2 V_1(n) +p_3 V_1(n-1) +p_4 V_1(n-2) +p_5 V_1(n-3)= 6(D-2)^3 J^3(1) J(n-2)\ ,
\end{equation}
\begin{equation*} 
\begin{split}
p_1=&-90n(n-1)(n-2)\ ,\quad p_2=9(n-1)(n-2)(16n-9D-6)\ ,\\
p_3=&(n-2)\left(-20n^2 + ( 122 - 21D )n  + 27D^2 - 66D - 42\right)\ ,\\
p_4=& -32n^3 + ( 54 + 93D )n^2  + ( - 78D^2 - 153D  +14)n + 18D^3 + 87D^2 + 21D - 24\ ,\\
p_5=& -(n-D-1)(n-2D+1)(2n-3D)\ . 
\end{split}
\end{equation*}
It contains in the r.h.s. the integral obtained from $V_1$ by 
contracting a line, which factorizes into a product of 4 one-loop tadpoles.
The solution of \eqref{eqv1} compatible with the large-$n$ boundary condition 
$V_1(n)\propto n^{-D/2}$ 
can be written as  $V_1(n)=C\; \IOMOG(n) +\INOMOG(n)\ $,
where $\IOMOG$ and $\INOMOG$ are respectively the 
solution of the homogeneous equation with above large-$n$ behaviour 
and a particular solution of 
the nonhomogeneous equation \itref{eqv1}. The constant $C$ turns out to be 
the value of the 3-loop vacuum bubble 
obtained from $V_1$ by deleting a line.

The amount of calculations needed to work out and solve all the systems of
difference equations is rather high, so that the calculations 
have been performed by means of an automatic tool, the program {\SYS }
described in Ref.\cite{Lapdif1}.

%.....................................................................%
%.....................................................................%

\eqrefb{V1}{V13} have a precision sufficient to substitute 
analytical expressions for most of practical uses.
Anyway, it is interesting to work out the analytical expressions of some  
of these results. In the case of $V_4$, a straightforward
application of the methods used in \cite{3-loop} for calculating 
the 3-loop $g$-$2$ master integrals gives 
\begin{multline}\label{V4anal}
V_4 =
  \dfrac{3}{2}\e^{-4}
 +\dfrac{19}{2}\e^{-3}
 +\dfrac{67}{2}\e^{-2}
 +\left(
 - 3\Z3 
 +\dfrac{127}{2} 
  \right)\e^{-1}
   - \dfrac{\pi^4}{20} 
   + 97 \Z3
   - \dfrac{237}{2}  
         + \e\biggl( 
	 189 \Z5 \\
	 -\dfrac{559}{60}\pi^4
	 - 32 \pi^2 \ln^2 2
	 + 768 \Li_4\left(\dfrac{1}{2}\right)
	 + 32 \ln^4 2 
	 + 1261 \Z3
	 -\dfrac{3969}{2}
	 \biggr)
+O(\e^2)\ ,
\end{multline}
which agrees perfectly with \eqref{V4}. 
As usual, the expression turns out to contain integer values of Riemann 
$\zeta$-function  $\zeta(p)=\sum_{n=1}^\oo 
1/{n^p}$, and values of the polylogarithms   
$\Li_p(x)=\sum_{n=1}^\oo {x^n}/{n^p}$. 
The same method may be also used for calculating $V_8$, but not for
calculating the remaining master integrals.
The reason is the appearance of double elliptic integrals 
which cannot be expressed in terms of polylogarithms. 

The appearance of elliptic integrals in the calculation of multiloop diagrams 
with many massive lines is a well-known phenomenon. 
In fact, the principal difficulty of the analytical calculation 
of 3-loop $g$-$2$ has been the individuation of the approaches which 
allowed us to avoid elliptic integrations. 
Unfortunately in the case of the 4-loop vacuum master integrals 
the problem cannot be circumvented and we are forced to perform the elliptic 
integrations.

Let us consider the simplest vacuum diagram, $V_1$.
By cutting one line of this bubble one obtains the 3-loop self-mass 
diagram shown in  Fig.2. If we take all masses equal to one, 
and the external line on the mass-shell, 
this diagram has two master integrals, whose $S_1$ is
\begin{equation*}
S_1=
\left({\pi^{D/2}\Gamma(1+\epsilon)}\right)^{-3}
  \int{d^D k_1 \; d^D k_2 \; d^D k_3}\ \dfrac{1}{Q}\ ,
\end{equation*} 
and $Q$ is the product of the four denominators of the diagram.

The analytical expressions of $V_1$ and $S_1$ are strictly related. 
In fact, introducing hyperspherical variables\cite{Levinehyp1,Levinehyp2} and
performing the angular integrations 
one finds that the finite parts of $V_1$ and $S_1$ can be expressed 
as a sum of many double elliptic integrals with the 
same three square roots in the 
denominator. The simplest of these integrals is that with unity as numerator
\begin{equation}\label{EllA}
A=\int^\oo_0 \dfrac{dl}{R(l,-1,-1)}
  \int^\oo_0 \dfrac{dm}{ R(m,l,-1) R (m,-1,-1)}= 2.641379476074689\ldots ,
\end{equation}
\begin{equation}
R(x,y,z)=\sqrt{x^2+y^2+z^2-2xy-2xz-2yz}\ .
\end{equation}
We have been not able to calculate this integral in analytical form.
Therefore, we have resorted to `experimental mathematics':
we have evaluated it at very high precision and we have tried
to fit the numerical value with various kinds of analytical  expressions.
Luckily\footnote{
%%%%%%%%%%%%%%%%%%%%%%%%%%%%%%%%%%%%%%%%%%%%%%%%%%%%%%%%%%%%%%%%%%%%%%%%%
We have been inspired by the analytical expressions of a double elliptic 
integral\cite{Prudnikov} related to the Appell $F_2$ function 
and by the analytical result of the 2-loop 3-lines ``sunrise'' 
self-mass diagram in two dimensions
\begin{equation*}\label{ell14}
S(p^2=-1,\  m_1^2=1,\  m_2^2=1-2^{-1/2},\  m_3^2=1+2^{-1/2},\  D=2)=
\pi K(1/2)/2=
\sqrt{\pi} {\Gamma^2(1/4)}/{8}\ .
\end{equation*}
},
%%%%%%%%%%%%%%%%%%%%%%%%%%%%%%%%%%%%%%%%%%%%%%%%%%%%%%%%%%%%%%%%%%%%%%%%%
we have found that
\begin{equation}\label{AEll}
A=K(w_{-})K(w_{+})\ , 
\qquad w_{\pm}=\dfrac{z_{\pm}}{z_{\pm}-1}\ ,
\qquad z_{\pm}=-(2-\sqrt{3})^4(4\pm\sqrt{15})^2\ ,
\end{equation}
where $K$ is the first of the two standard elliptic integrals 
\begin{equation}
K(m)=\int^1_0 \dfrac{dt}{\sqrt{1-t^2}\sqrt{1-m t^2}}\ , 
\qquad 
E(m)=\int^1_0 \dfrac{dt\;\sqrt{1-m t^2}}{\sqrt{1-t^2}}\ .
\end{equation}
We have verified \eqref{AEll} with a precision of 
more than 30000 digits\footnote{
%%%%%%%%%%%%%%%%%%%%%%%%%%%%%%%%%%%%%%%%%%%%%%%%%%%%%%%%%%%%%%%%%%%%%%%%%
By making the replacement $R(l,-1,-1)$ $\to$ $R(l,-1,\lambda)$, 
working out the differential equation in $\lambda$ satisfied by $A(\lambda)$,
and evaluating $A(1)$ from the series expansion of the solution, with the initial
condition $A(0)=\pi^2/4$.
}
%%%%%%%%%%%%%%%%%%%%%%%%%%%%%%%%%%%%%%%%%%%%%%%%%%%%%%%%%%%%%%%%%%%%%%%%%
. 
Once the analytical expression of the basic integral $A$ has been 
identified beyond any reasonable doubt,
we have considered the full integral representations
 of the finite parts of $V_1$ and $S_1$;
 they contain integrals which differ from \eqref{EllA} by 
additional polynomials and logarithmic functions in the numerators.
A high-precision numerical expansion in $\e$ 
of $S_1$ was calculated in Ref.\cite{Lapdif1a} 
by means of the program {\SYS }. 
We have been able to fit 
the numerical result
with the analytical expression
%\footnote{
%%%%%%%%%%%%%%%%%%%%%%%%%%%%%%%%%%%%%%%%%%%%%%%%%%%%%%%%%%%%%%%%%%%%%%%%%
%In Ref.\cite{Berends} the master integral $S$ was expressed in terms 
%of Lauricella hypergeometric functions of several variables. 
%Unfortunately, analytical expressions of the values of these multiple series
%%like \eqref{CEll} 
%are not known.
%}
%%%%%%%%%%%%%%%%%%%%%%%%%%%%%%%%%%%%%%%%%%%%%%%%%%%%%%%%%%%%%%%%%%%%%%%%%

\begin{multline}\label{CEll}
S_1(\text{Fig.2})=
 2\e^{-3}+\dfrac{22}{3}\e^{-2} + \dfrac{577}{36}\e^{-1}
+\dfrac{4\pi}{\sqrt{15}}
  \left(
 \dfrac{35}{8}\pi +\dfrac{131}{12}K(w_{-})K(w_{+})
  \right.\\\left.
 -\dfrac{7}{2}
 \left(E(1-w_{-})E(1-w_{+})+5E(w_{-})E(w_{+})\right)
 \right) 
 +\dfrac{6191}{216} +O(\e)\ 
\end{multline} 
by using the integer-relation search algorithm PSLQ\cite{PSLQ}.
We have verified \eqref{CEll} with 1200 digits of precision.
\eqref{CEll} is important because it shows for the first time that there exist
single-scale integrals which contain non-polylogarithmic transcendental
constants. The same constants also appear in the analytical expression of 
$V_1$, together with higher-transcendentality constants so far unidentified.

This result also sheds light on the possible analytical structure of  
of the 4-loop QED contribution to the electron $g$-$2$. In fact 
one of the 4-loop $g$-$2$ master integrals is a  
4-loop 5-lines self-mass integral analogous 
to $S_1$, expressible in terms of irreducible triple elliptic integrals 
%containing 4 square roots 
analogous to \eqref{EllA}.

%%%%%%%%%%%%%%%%%%%%%%%%%%%%%%%%%%%%%%%%%%%%%%%%%%%%%%%%%%%%%%%%%%%%%%%%%%%%%%
%%%%%%%%%%%%%%%%%%%%%%%%%%%%%%%%%%%%%%%%%%%%%%%%%%%%%%%%%%%%%%%%%%%%%%%%%%%%%%

%%%%%%%%%%%%%%%%%%%%%%%%%%%%%%%%%%%%%%%%%%%%%%%%%%%%%%%%%%%%%%%%%%%%%%%%
\vfill\eject 
\pagenumbering{roman}
\setcounter{page}{1}
\phantom{.}\vspace{5cm}
\section*{Figure Captions}
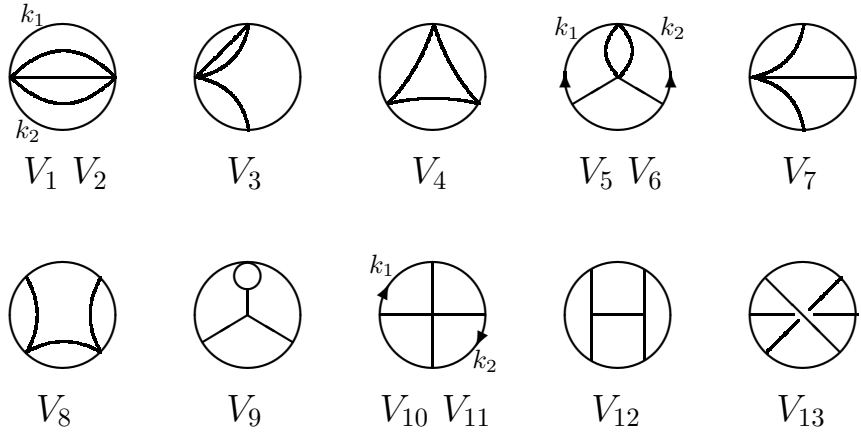
\par\noindent Figure 1: \CAPTIONFIGA
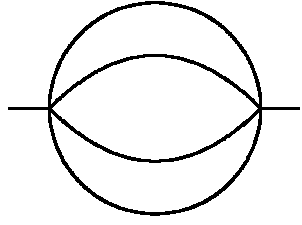
\par\noindent Figure 2: \CAPTIONFIGB
\phantom{.}\vspace{12cm}\IDENTIFY
%%%%%%%%%%%%%%%%%%%%%%%%%%%%%%%%%%%%%%%%%%%%%%%%%%%%%%%%%%%%%%%%%%%%%%%%
\phantom{.}
\vfill\eject 
\phantom{.}\vspace{1cm}
\FigureOne
\phantom{.}\vspace{1cm}\IDENTIFY
%%%%%%%%%%%%%%%%%%%%%%%%%%%%%%%%%%%%%%%%%%%%%%%%%%%%%%%%%%%%%%%%%%%%%%%%
\phantom{.}
\vfill\eject 
\phantom{.}\vspace{1cm}
\FigureTwo
\phantom{.}\vspace{1cm}\IDENTIFY
%%%%%%%%%%%%%%%%%%%%%%%%%%%%%%%%%%%%%%%%%%%%%%%%%%%%%%%%%%%%%%%%%%%%%%%%
%%%%%%%%%%%%%%%%%%%%%%%%%%%%%%%%%%%%%%%%%%%%%%%%%%%%%%%%%%%%%%%%%%%%%%%%
%%%%%%%%%%%%%%%%%%%%%%%%%%%%%%%%%%%%%%%%%%%%%%%%%%%%%%%%%%%%%%%%%%%%%%%%
%%%%%%%%%%%%%%%%%%%%%%%%%%%%%%%%%%%%%%%%%%%%%%%%%%%%%%%%%%%%%%%%%%%%%%%%
%%%%%%%%%%%%%%%%%%%%%%%%%%%%%%%%%%%%%%%%%%%%%%%%%%%%%%%%%%%%%%%%%%%%%%%%
%%%%%%%%%%%%%%%%%%%%%%%%%%%%%%%%%%%%%%%%%%%%%%%%%%%%%%%%%%%%%%%%%%%%%%%%
%%%%%%%%%%%%%%%%%%%%%%%%%%%%%%%%%%%%%%%%%%%%%%%%%%%%%%%%%%%%%%%%%%%%%%%%
%%%%%%%%%%%%%%%%%%%%%%%%%%%%%%%%%%%%%%%%%%%%%%%%%%%%%%%%%%%%%%%%%%%%%%%%
%%%%%%%%%%%%%%%%%%%%%%%%%%%%%%%%%%%%%%%%%%%%%%%%%%%%%%%%%%%%%%%%%%%%%%%%%%%%%%
\end{document}